# DETECTION OF CALENDAR-BASED PERIODICITIES OF INTERVAL-BASED TEMPORAL PATTERNS

Mala Dutta[1] and Anjana Kakoti Mahanta[2]

[1]Department of Computer Science, Gauhati University, Guwahati, Assam
maladuttasid@gmail.com
[2]Department of Computer Science, Gauhati University, Guwahati, Assam
anjanagu@yahoo.com

## ABSTRACT

*We present a novel technique to identify calendar-based (annual, monthly and daily) periodicities of an interval-based temporal pattern. An interval-based temporal pattern is a pattern that occurs across a time-interval, then disappears for some time, again recurs across another time-interval and so on and so forth. Given the sequence of time-intervals in which an interval-based temporal pattern has occurred, we propose a method for identifying the extent to which the pattern is periodic with respect to a calendar cycle. In comparison to previous work, our method is asymptotically faster. We also show an interesting relationship between periodicities across different levels of any hierarchical timestamp (year/month/day, hour/minute/second etc.).*

## 1. INTRODUCTION

Identifying and extracting patterns and regularities in massive data repositories has been a focused theme in data mining research for almost over a decade. Substantial progress continues to be made in this context, specially in the tasks of frequent itemset mining [15] and association rule mining [7], [8], [10], [16], [17]. Temporal pattern discovery is a very promising extension to this ongoing research theme because it substantially broadens the scope of data analysis by supporting the discovery of patterns and regularities that are time-dependent. For example, across a time-series, a domain-specific pattern may occur only in some time-periods. Say, in a temperature time-series, a 10°F temperature rise could occur in certain time-periods. In a sales data archive, panic reversal of sales could occur in certain time-intervals. In data streams of stock prices, the prices may rise(peak) in some time-periods and so on. It is often useful to know if a pattern is periodic or not. To determine periodicities of a domain-specific pattern that occurs in some time-periods across a time-series, at first all these time-periods have to be extracted from the time-series and then the nature of this sequence of time-intervals has to be studied. Similarly, periodicities of a natural event such as volcanic eruption, tropical storm across a particular region etc can be determined from an event-related dataset which records the time-intervals in which the event has occurred. In the underlying context, even an event will simply be referred to as a pattern. In this paper, a technique is proposed to extract calendar-based periodicities – viz. yearly periodicities, monthly periodicities etc of an interval-based temporal pattern i.e. of a pattern that occurs across a sequence of time-intervals in either a discrete or in a continuous domain. Some preliminary work in this area for patterns in a discrete domain only was done by Dutta and Mahanta [2]. In our paper, a function called occurrence function is defined for a timestamp. Given a sequence of time-intervals in either a discrete or a continuous domain, a generalized algorithm has been developed for computing the occurrence function value at any timestamp. Next, an algorithm for locating local maxima of the occurrence function is also proposed. The correctness of both the proposed algorithms is established mathematically. These two algorithms are then used to extract calendar-based periodicities of patterns that occur across a sequence of time-intervals in a discrete or in a continuous domain. The proposed method to extract calendar-based periodicities can be used to extract both partial as well as full periodicities of interval-based temporal patterns with the same efficiency. The extraction of periodicities takes O($n log n$) time for a continuous domain and O($n$) for a discrete domain. Finally in this paper, a theorem that captures a relationship between the periodicities of patterns at different levels of any time-hierarchy is also formulated and subsequently proved.

The paper has been organized as follows - some recent works done in mining of periodicities of patterns in temporal data are mentioned in section 2. Section 3 gives a brief description of the classic dynamic time-warping (DTW) technique that has been used here to extract the time-intervals in which a domain-specific pattern appears across a time-series. The definition of the occurrence function and the algorithms to compute the function and to locate local maxima of the function are presented in section 4. In section 5, it is shown how the algorithms proposed in section 4 can be used for extracting calendar-based (i.e. seasonal) periodicities of interval-based temporal patterns in a discrete or in a continuous domain. A theorem establishing a relationship between the periodicities of patterns at different levels of a time-hierarchy is presented in section 6. The results obtained after applying the proposed technique on real-life datasets are given in section 7. Section 8 gives the conclusion and mentions the scope for further research in this line.

## 2. RECENT WORKS DONE IN THIS FIELD

Mining periodicities of patterns in temporal data is an active research area. Some recent works in this field are mentioned below:

Elfekey et al. [5] propose algorithms to mine two pre-defined types of periodicities in time-series data. Berberidis et al. [1] propose an algorithm that mines a set of candidate periods featured in a time-series that satisfy a minimum confidence threshold. Elfekey et al. [3] propose an algorithm for mining periodic patterns in time-series databases with unknown or obscure periods.Yang et al. [18] and Huang and Chang [6] propose algorithms for mining asynchronous periodic patterns in time-series data. Lai et al. [11] address the problem of mining periodicity of patterns that occur across artificial boundaries. Karli and Saygin [9] propose two techniques for mining periodic spatio-temporal patterns at different time granularities. Zhang et al. [19] present practical algorithms to solve the problem of mining frequently occurring periodic patterns with a gap requirement from sequences. Ma and Hellerstein [13] study partial periodic patterns taking into account imprecise time information, noisy data and shifts in phase and/or periods. Algorithms for incremental mining of partial periodic patterns in time-series archives are proposed and analyzed empirically by Elfekey et al. [4]. Lee et al. [12] address the problem of mining multiple partial periodic patterns in a parallel computing environment. Mahanta et al. [14] and Dutta and Mahanta [2] propose algorithms to extract calendar-based periodic temporal patterns across discrete domains. In this paper, a generalized method is proposed to detect calendar-based periodicities of temporal patterns occurring across a sequence of time-intervals not only in a discrete domain but across a continuous domain also. A theorem establishing an interesting relationship between periodicities of patterns at different levels of a time-hierarchy is also formulated.

## 3. DYNAMIC TIME-WARPING METHOD

For the automatic detection of a pattern in a time-series, an approximate or "fuzzy" matching process is required that can capture all the time-series fragments within which the approximate shape of the pattern is detected. Specifically, the pattern detection task involves searching a time-series $S = s_1, s_2, s_3 ...s_n$ for instances of a given template $T = t_1, t_2, t_3 ...t_m$. In the dynamic time-warping (DTW) technique, this pattern detection task is achieved by applying a dynamic programming approach to align the two sequences $S$ and $T$ in a way so that some distance measure is minimized. To achieve a reasonable fit, the time series may be stretched or compressed. Shown below in Figure 1 is a n-by-m grid where each grid-point (i,j) corresponds to an alignment between elements $s_i$ and $t_j$ of $S$ and $T$ respectively.

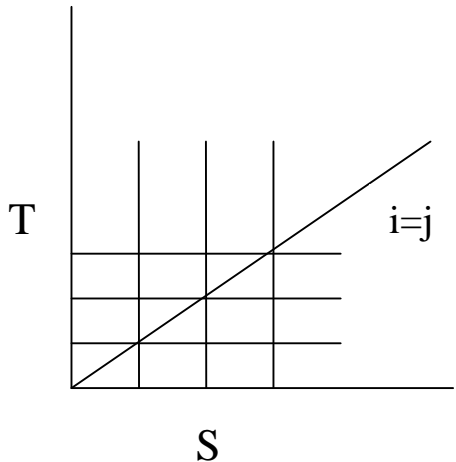

Figure 1

A warping path W is a path through this grid that aligns the elements of S and T such that the distance between them is minimized. The warping path W is thus represented as $W = w_1 w_2 ...........w_p$ where each $w_k$ corresponds to a point $(i,j)_k$ in the grid. The dynamic time-warping problem can hence be formally defined as a minimization over potential warping paths based on the cumulative distance for each path.

$$DTW(S,T) = \min_W \{ \sum_{k=1}^{p} \delta(w_k) \}$$

where $\delta$ is the distance measure used to compute the distance between any two elements $s_i$ and $t_j$. Typically, $\delta(i, j) = |s_i - t_j|$ is used. Searching through all possible warping paths leads to a combinatorial explosion. Several restrictions are hence placed on permissible paths between two grid points, thereby reducing the search space – viz. all possible warping paths. A few restrictions for the warping path are outlined below:

Monotonicity: The points in W are monotonically ordered with respect to time i.e. for consecutive points – $w_k$-1 and $w_k$ in W, $i_k$-1 $\leq$ $i_k$ and $j_k$-1 $\leq$ $j_k$.
Continuity: The allowable steps taken by the path W in the grid is confined to neighboring points i.e. $i_k - i_k$-1 $\leq$ 1 and $j_k - j_k$-1 $\leq$ 1.

Warping Window: Allowable points are constrained to fall within a given warping window,| $i_k - j_k$ | ≤ w, where w is the size of the warping window which is a positive integer.

The dynamic programming formulation is based on the following recurrence relation which defines the cumulative distance (i ,j) for each point.

(i, j) =  (i, j) + min[ (i - 1, j),  (i - 1, j - 1),  (i, j - 1)]

i.e. The cumulative distance is the sum of the distance between current elements (specified by a grid point) and the minimum of the cumulative distances of the neighboring points. The cumulative distance associated with the best warping path is simply a raw score. Normalization of the raw score is necessary to ensure that matches differing only in scale are comparable. Normalization is also required to accommodate differences in path length i.e. the number of grid points in a warping path. The normalized cumulative distance is used to determine if the degree of fit of the time-series S and the given template T is sufficiently good or not .

# 4. Computation of the occurrence function and detection of local maxima of the function

As mentioned earlier, a sequence of time-intervals is associated with an interval-based temporal pattern signifying the time-periods in which that pattern occurs. Assuming that there exists a sequence of n time-intervals and t is a timestamp that appears in m (m   n) number of these n time-intervals, then the value of the occurrence function (t) at t is defined to be m. If the intervals are all disjoint, then the value of the occurrence function at any time-stamp will be either zero or one.

Two theorems are presented below that establish some properties of the occurrence function. On the basis of these properties, changes of the occurrence function will be identified and this information will subsequently be used for computing the value of the function at any timestamp *s* and for detecting local maxima of the function.

## 4.1. Properties of the occurrence function

The following two theorems valid for continuous domains  state some interesting properties of the occurrence function :

Theorem 1  Let the endpoints of a given sequence of time-intervals be merged into a single list and this list (of endpoints) is then sorted in ascending order of timestamps. Let a and b be the timestamps of two successive endpoints with a<b in the sorted list of endpoints. Then the occurrence function is a constant in time interval (a,b).

Proof: - Let  x, y   (a,b). Let x belong to one of the given intervals which will be of the form either [c,d]  or (c,d)  or (c,d]  or  [c,d). Since there is no endpoint in (a,b) and x   d and d is an endpoint, we must  have  b  d. Similarly c  a. Therefore y also belongs to the same interval with endpoints c and d. Similarly if y belongs to one of the given intervals then, x also belongs to the same interval. Therefore the number of given intervals containing x and the number of intervals containing y are same and hence  (x) =  (y). This proves the theorem.

Theorem 2  Let t be a timestamp where some endpoint occurs and let

L  = Lim $_{x -> t-}$  (x)

R $= \mathrm{Lim}_{x->t+}$ (x)

$n_1$ = number of left open endpoints at t

$n_2$ = number of left closed endpoints at t

$n_3$ = number of right open endpoints at t

and $n_4$ = number of right closed endpoints at t

Then (t) $- L = n_2 - n_3$          (1)

and R $- L = n_1 + n_2 - n_3 - n_4$          (2)

Proof :- Let y be a point between t and the timestamp at which an endpoint occurs just before t. Let k be the number of given intervals containing y. Obviously L = k since the occurrence function remains constant for all such *y* (Theorem 1). Out of these k intervals, $n_3$ of them will have right open endpoints at t and thus t will no longer be contained in these $n_3$ intervals. However there are $n_2$ left closed endpoints at t which means that $n_2$ new intervals will start containing t. Therefore (t) $= k + n_2 - n_3 = L + n_2 - n_3$ and hence (t) $- L = n_2 - n_3$. If t is the timestamp for the first endpoint, $n_3$ and L must be zero and $n_2$ of the given intervals will contain t. Therefore again, (t) $- L = n_2 - n_3$.This proves (1). Similarly, we can prove (t) $- R = n_4 - n_1$. Subtracting this equation from (1), we get R $- L = n_2 - n_3 + n_1 - n_4$ which proves (2). This completes the proof of the theorem.

A discrete domain can be extended to a continuous one by inserting all the points between two successive endpoints. Thus the theorems given above that are valid for a continuous domain can be used for discrete domains also.

## 4.2. Capturing the changes of the occurrence function

On the basis of the theorems presented above in section 4.1, we now propose a method to identify changes of the occurrence function. This information will be used for finding the function value at any timestamp *s* (Section 4.3) and for detecting local maxima of the function (Section 4.4).

To identify changes of the occurrence function, all the endpoints of an existing sequence of n time-intervals are first merged into a single list. An endpoint record has two fields - a timestamp t and et giving the endpoint type (left open, right open, left closed or right closed). After an existing sequence of n time-intervals is converted to a list of endpoint records, the list of endpoint records is then sorted in ascending order of timestamps. The changes of the occurrence function can be identified by simply scanning a sorted list of endpoint records. Information about each change of the occurrence function is captured in a change record. A change record will have three fields – timestamp t at which a change of the occurrence function is observed, u which gives the value of the occurrence function at t and r which is the right hand limit of the occurrence function at t for a continuous domain or the value of the occurrence function at the next timestamp for a discrete domain.

The algorithm given below (Algorithm 1) creates an array d of change records from an existing array e of endpoint records (assumed to be sorted in ascending order of timestamps). The length of the array e is 2n, where n is the total number of time-intervals. The length of the array d will be m which is the number of changes of the occurrence function. The value of m will be less than or equal to the number of endpoints with distinct timestamps. Obviously m  2n. Scanning the array e of sorted endpoint records in ascending order of timestamps, the values of $n_1$, $n_2$, $n_3$ and $n_4$ (as defined in Section 4.1) are counted for each timestamp t in the array. Theorem 2 is used to identify each change of the occurrence function and for each change, an element is added to the array d with the fields - timestamp t, u = $n_2 - n_3 + $ L and r = $n_1 + n_2 - n_3 - n_4 + $ L Here L is the left-hand limit of the occurrence function at t for a continuous domain or for the continuous extension of a discrete domain. The value of L is equal to the value of r of the previous change (by Theorem 1). At this point, d[i].t and d[i].u give the timestamp and the value of the occurrence function at the $i^{th}$ change of the occurrence function. These values are correct for both continuous and discrete domains. Also d[i].r correctly gives the right hand limit of the occurrence function at d[i].t for a continuous domain. For a discrete domain, d[i].r at this point gives the right hand limit of the occurrence function at d[i].t for the continuous extension of the discrete domain. For a discrete domain, as mentioned earlier, d[i].r is supposed to give the value of the occurrence function at the next timestamp after d[i].t. So in a discrete domain, if d[i+1].t is greater than d[i].t + 1 then d[i].r correctly gives the desired value of the occurrence function at d[i].t + 1 (by Theorem 1). But if d[i+1].t = d[i].t + 1, then d[i].r needs to be replaced by d[i+1].u .These replacements are done in the last part of Algorithm 1.

```
Algorithm 1:

     i ← 1
     m = 0
     L = 0

     while (i <= 2n)
       {
           n1 = n2 = n3 = n4 = 0
           ct ← e[i].t
           while (i <= 2n  &&  ct equal  to e[i].t)
           {
             increment n1 or n2 or n3 or n4 according to e[i].type
             i++
           }

           if ( ( ( n2 − n3 )  ≠  0)  or ( (n1 + n2 − n3 − n4) ≠ 0 ) )
           {  m++
             d[m].t =  ct
             d[m].u = n2 - n3 + L
             d[m]. r = n1 + n2 − n3 − n4 + L
             L = d[m].r
           }
       }

/* Do the following for discrete domain */

     for i =  1  to  m - 1
        if  d[i+1].t = = d[i].t + 1
           d[i].r =  d[i+1].u
```

### 4.3. Finding the occurrence function value at any timestamp s

After capturing the changes of the occurrence function for a given sequence of n time-intervals as described above in Section 4.2, the occurrence function value at any timestamp s can be calculated by a binary search on the array d of change records (created by Algorithm 1) of length m sorted in timestamp order.

If $s = d[i].t$ for some i, then the value of the occurrence function at s is $d[i].u$. If s $< d[1].t$ or $s > d[m].t$ then the value of the occurrence function at s is zero. Finally if $d[i].t < s < d[i+1].t$ then the value of the occurrence function at s is $d[i].r$. The correctness of this for continuous domains follows from Theorem 1. For discrete domains this condition implies that $d[i+1].t > d[i].t + 1$ and by Algorithm 1, $d[i].r$ is the right hand limit of the occurrence function in the continuous extension of the discrete domain which correctly represents the value of the occurrence function at s (by Theorem 1).

### 4.4. Detection of local maximal of the occurrence function

The change records described in Section 4.2 actually record two changes of the occurrence function – (i) the change from the left hand limit to the value of the function at a timestamp t and (ii) the change from the value of the function at that timestamp t to the right hand limit. For the detection of local maxima of the occurrence function, it is however convenient to introduce a knot record that will capture only a single change in the value of the occurrence function. A knot record has only two fields – timestamp t and v which gives the value of the occurrence function at t. For obtaining information about local maxima of the occurrence function, at first an array k of knot records has to be created from an array d of change records. Traversing the array d in ascending order of timestamps, for each $i^{th}$ (1 i m) change record $d[i]$, at most two knot records are added to the array k in the following order - A knot record with fields v set to $d[i].u$ and t set to $d[i].t$ is added followed by another knot record with fields v set to r and t set to $d[i].t$ for the continuous case and $d[i].t + 1$ for the discrete case. A knot record is however not added to the array k if the value of it's v field is same as that of the just previously added knot record in the array k. Adding the knot records to the array k in the order as explained above automatically ensures that the knot records in the array k are arranged in non-decreasing order of timestamps. Now if $(t_1,v_1)$ and $(t_2,v_2)$ are two successive knot records in the array k and there exists a timestamp t such that $t_1 < t < t_2$, then the value of the occurrence function at t will be $v_1$. This is because each knot record captures one change in the occurrence function value and the occurrence function is a constant between two successive points of change.

The algorithm given below (Algorithm 2) identifies local maxima of the occurrence function by traversing an array k of knot records. Let p be the length of this array k. Information about each local maximum is captured in a lmaxnode record having the following fields - timestamps start, peakstart, peakend, end and occurrence function values startval, peakval and endval.

Algorithm 2:

state = increasing
set current to a  new lmaxnode record
current.start = k[1].t  for continuous  domain
                and  k[1].t - 1  for  discrete domain
current.startval = 0
for i = 1 to p
   { if  state = = increasing  and  k[i+1].v < k[i].v    (see Figure 2)
        { current.peakstart = k[i].t
           current.peakend = k[i+1].t  for  continuous domain
                             and  k[i+1].t-1 for  discrete domain
           current.peakval = k[i].v
           state = decreasing
         }
     if  state = = decreasing and (( i = = p) or ( k[i+1].v > k[i].v ))  (see Figure3)
        { current.end =  k[i].t
           current.endval = k[i].v
           add current to lmaxlst
           if  i < p
           { set  current to a new lmaxnode
             current.start = k[i+1].t  for continuous  domain
                     and  k[i+1].t – 1  for  discrete  domain
             current.startval = k[i].v

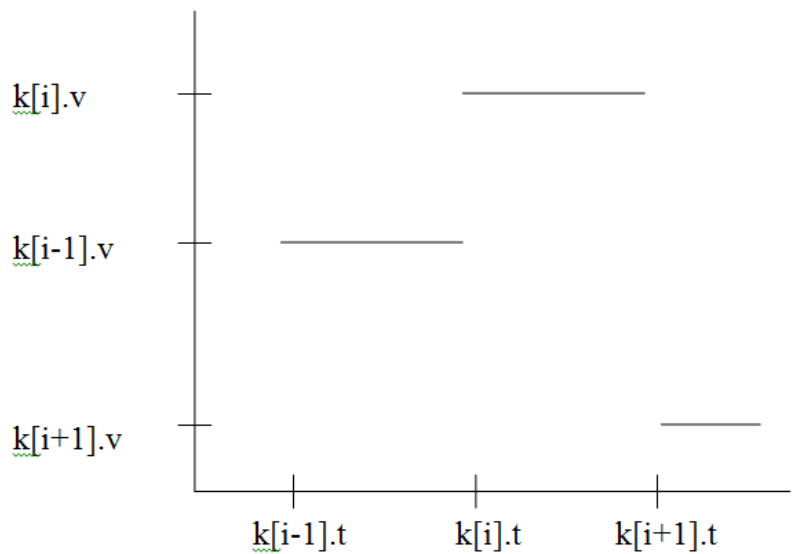

Figure 2   state  =  increasing  and   k[i +1].v < k[i].v  for  continuous  domain

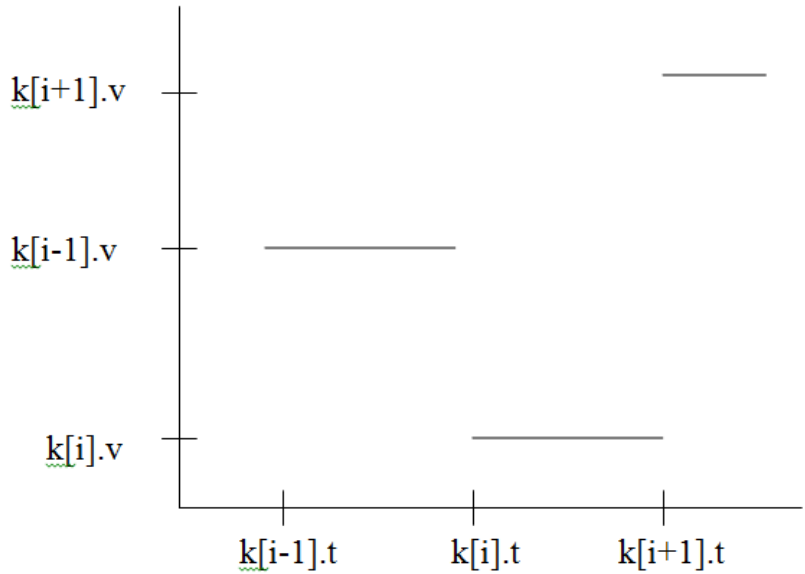

Figure 3. state = decreasing and k[i +1].v > k[i].v for continuous domain

## 4.5. Time Complexity

Creating the array e of endpoint records takes O(n) time where n is the number of input intervals. Sorting this array e of endpoint records will take O(n log n) time in the worst-case for continuous domains. For discrete domains, if the timestamps are hierarchical with each field having a limited range, radix sort can be used which performs in O(n) time. Scanning the array e for changes of the occurrence function and then creating the array d of change records will take O(n) time in the worst-case. Creating the array k of knot records from the array d will take O(n) time in the worst-case. Scanning for local maxima in the array k of knot records takes O(n) time. Thus the overall worst-case time-complexity works out to be O(n log n) for continuous domains and O(n) for discrete domains (for hierarchical timestamps). Moreover, once the above mentioned operations are performed and the corresponding structure is set up, to compute the occurrence function at any arbitrary point takes only O(log n) time in the worst-case.

## 5. DETECTION OF CALENDAR-BASED PERIODICITIES OF PATTERNS

In the underlying context, a pattern will be characterized by a sequence of time-intervals in which the pattern occurs. A pattern here typically refers to a natural event such as a hurricane, volcanic eruption, outbreak of malaria etc. It may also refer to a domain-specific pattern such as panic reversal of sales, a temperature trend, rise (peak) in stock prices etc. We are interested in determining if a pattern is periodic. Assuming that the timestamps associated with a pattern are calendar-dates (i.e. of the format day-month-year or time-day-month-year etc.), we propose a method to extract calendar-based periodicities viz. yearly periodicities, monthly periodicities etc of the pattern using the algorithms described in Section 4. If a pattern is seen every year in the month of April say, then we call it a yearly pattern. Similarly suppose a pattern is seen in the first week of every month, then we call it a monthly pattern and so on.

The time-intervals in which a pattern occurs are maintained in a list L. If any pattern occurrence spans into different year(s), then that time-period has to be split up into two or more corresponding time-intervals and only these are to be inserted into the list L. A few examples are shown below in Table 1.

Table 1. Handling pattern occurrences spanning across different year(s)

| Pattern Occurrence | Corresponding time-intervals to be inserted into the list L |
|---|---|
| 18th Dec, 2001 to 7th Jan, 2002 | 18th Dec, 2001 to 31st Dec, 2001<br>1st Jan, 2002 to 7th Jan, 2002 |
| 24th Dec, 2005 to 15th Jan, 2007 | 24th Dec, 2005 to 31st Dec, 2005<br>1st Jan, 2006 to 31st Dec, 2006<br>1st Jan, 2007 to 15th Jan, 2007 |
| 28th Dec, 2009 to 1st Jan, 2010 | 28th Dec, 2009 to 31st Dec, 2009<br>1st Jan, 2010 to 1st Jan, 2010 |

It is also to be noted that the time-intervals in the list L have to be disjoint. If the pattern has occurred in overlapping time-periods, then the overlapping periods are merged and only the merged time-period is inserted into the list L. E.g. if a pattern occurs from 10th June, 2001 to 18th June, 2001 and again from 14th June, 2001 to 20th June, 2001, then only the corresponding merged time-interval (10th June, 2001 to 20th June, 2001) will be inserted into the list L. The necessity of keeping the time-intervals in the list L disjoint shall be explained later in this unit.

To extract calendar-based periodicities of the pattern, timestamps of the disjoint time-intervals in list L are stripped of certain components. While finding yearly periodicities, the year component of the timestamps is removed from the corresponding dates. Similarly when monthly periodicities are being searched, the year and month components are not considered. Again while looking for daily periodicities, the day, month and year components are stripped and so on. Let L´ be a list containing all the time-intervals in the list L but with timestamps i.e. calendar dates stripped of the appropriate component(s). L´ now will possibly have overlapping intervals. Now with this list L´ as the input, an array d of occurrence function change records (as described in Section 4.2) is created using Algorithm 1 (which is also described in Section 4.2). Next from this array d, an array k of knot records (as described in Section 4.4) is created. Using Algorithm 2 (also described in Section 4.4), the array k of knot records is scanned to extract local maxima of the occurrence function values. It is easy to see that each local maximum gives a periodicity of the pattern under study. It is possible to distinguish between partial and full periodicities of a pattern. We elaborate on this below-If s is the smallest timestamp and g is the largest timestamp appearing in the time-intervals of the list L, then the period from s to g is called the lifespan of the pattern in the underlying context. A certainty function c(x) is defined for a stripped timestamp x of the list L´ as

$$c(x) = \rho(x) \big/ N$$

Here (x) is the occurrence function value for the stripped timestamp x and N is the total number of periods in the lifespan of the pattern. N gives the total number of years or months or days or hours etc. depending on whether we are looking for yearly periodicities or monthly periodicities or daily periodicities or hourly periodicities etc. The maximum value of c(x) will be 1 because the time-intervals in the list L (from which list L´ was created) are all disjoint. Within a particular identified local maximum of the occurrence function, if the certainty function c(x) reaches it's maximum value i.e. 1 at any timestamp x, then the pattern is fully periodic at x. On the other hand, within a particular identified local maximum of the occurrence function, if the certainty function $c(x)$ is less than 1 at any timestamp x, then the pattern will be partially periodic at x.

For example, suppose a pattern with a lifespan of ten years is scanned for yearly periodicities and a local maximum (of the occurrence function) is detected from $10^{th}$ July to $15^{th}$ July. Now if the pattern is observed on $12^{th}$ July every year in all ten years, then the value of the certainty function on $12^{th}$ July becomes equal to 1 and this yields a full yearly periodicity of the pattern. On the other hand, if the pattern is seen on $12^{th}$ July only in eight out of ten years, then the certainty function value on $12^{th}$ July will be 0.8, thereby yielding a partial yearly periodicity of the pattern.

The necessity of keeping the time-intervals in the list L all disjoint is now explained in the context of the example given above. Let us assume that the time-intervals in the list L are not disjoint. Now, suppose the pattern is noticed on $12^{th}$ July in ten overlapping time-periods in a particular year but not even once in any of the remaining nine years. Though in this case, the pattern is obviously not periodic at all on $12^{th}$ July, still the value of the certainty function on $12^{th}$ July becomes equal to 1. For this reason, to be able to determine periodicities properly, all the time-intervals in the list L necessarily need to be disjoint.

## 6. A RELATIONSHIP BETWEEN PERIODICITY OF PATTERNS AT DIFFERENT LEVELS OF A TIME-HIERARCHY

The timestamps associated with temporal patterns often have a hierarchical structure. In the previous section (Section 5), a method was proposed for detecting calendar-based periodicities of patterns whose timestamps had a calendar-date based time-hierarchy. In this section, a theorem is presented that shows an interesting relationship between the periodicity of patterns at different levels of a time-hierarchy.

Theorem 3 If $l_k$ is the $k^{th}$ level of the time-hierarchy (starting at the lowest level) and if for $i < j$, there are p combinations of values of levels $l_i$, $l_{i+1}$ ....$l_{j-1}$ for every $l_j$ value, then a level $l_i$ periodic pattern of periodicity $f > (p-1)/p$ will give rise to p level $l_j$ periodic patterns of average periodicity f and minimum periodicity $1 - p(1- f)$.

Proof :- Let the level $l_i$ pattern occur at a date d obtained after levels $l_k$ for k ≥ i are stripped . Suppose in the lifespan, there are $m_i$ periods for level $l_i$ and $m_j$ periods for level $l_j$ . Then obviously

$$m_i = p * m_j \qquad (1)$$

Also the pattern occurs at date d for $f * m_i$ level $l_i$ periods. Since there are p combinations $c_1$, $c_2$, ....,$c_p$ of values of levels $l_i$, $l_{i+1}$ ......$l_{j-1}$ for every $l_j$ value, the pattern appears at dates $(c_1,d)$, $(c_2,d)$,.....$(c_p,d)$ (after stripping levels $l_k$ for k ≥ j) for some $n_1$, $n_2$, ....,$n_p$ level $l_j$ periods respectively where

$$n_1 + n_2 + ......+ n_p = f * m_i \qquad (2)$$

Since $n_k$ ≤ $m_j$, $n_k$ ≥ -(p-1) $m_j$ + f * $m_i$

$$= [-(p-1) + fp] m_j$$

Hence $\quad n_k \quad m_j [fp - (p-1)]$

$$> m_j [(p-1) - (p-1)] \;=\; 0$$

Thus there are p level $l_j$ periodic patterns at the dates $c_1.d, c_2.d, \ldots\ldots c_p.d$ of minimum periodicity Min $(n_k / m_j) = fp - (p-1) = 1 - p(1-f)$ and with average periodicity

$$_k \,(n_k / m_j) / p$$

$$= (\,_k\; n_k) / (m_j * p)$$

$$= (f * m_i) / (m_j * p) \qquad\qquad \text{using (2)}$$

$$= (f * m_i) / m_i \qquad\qquad\qquad \text{using (1)}$$

$$= f$$

This proves the theorem.

We note that if f is 1 i.e. if the level $l_i$ pattern is fully periodic, then each of the level $l_j$ periodic patterns has periodicity 1 and hence is 1 i.e. is fully periodic. Because of this theorem, a fully periodic monthly pattern will give rise to 12 fully periodic yearly patterns, a fully periodic daily pattern will give rise to 365 fully periodic yearly patterns and so on.

For an example of a partially periodic pattern, let us consider a monthly pattern of periodicity f = 23/24 on the $10^{th}$ of a month. This monthly pattern will give rise to 12 (p =12) yearly patterns at dates Jan 10, Feb 10, … , Dec 10. Let the lifespan of the data be 10 years i.e. $m_j = 10$ and hence $m_i = 120$. Thus the pattern appears on the $10^{th}$ in (23*120) / 24 = 115 months. Let $n_k$ of these be in month k (k = 1 for Jan, k = 2 for Feb etc). Then $_k\, n_k = f * m_i = 115$. Since the sum is constant, any one of them will be minimum if the rest of them are maximum. Now the maximum of any $n_k$ is 10 (i.e. $m_j$) since month k can appear at most for 10 years. Therefore minimum of any $n_k = 115 – (11 * 10) = 5$. Thus the minimum periodicity is $5/10 = \frac{1}{2}$ which is $1 – p(1 – f)$ as given by the theorem. The sum of the periodicities of these 12 patterns is $115/10 = 23/2$. Hence the average is $(23/2)/12 = 23/24 = f$ as given by the theorem.

Though the above examples have used a calendar-date based time-hierarchy, Theorem 3 holds true for any other time-hierarchy also.

## 7. EXPERIMENTAL RESULTS

The technique proposed in Section 5 for extracting calendar-based periodicities of interval-based temporal patterns has been applied here to an event-related dataset and then to a time-series.

Working with an event-related dataset : The dataset contains the time-periods of occurrence of a category 1 hurricane (category based on Saffir-Simpson scale) across the eastern-pacific region. The source of this data is http://weather.unisys.com/hurricane/index.html. The technique proposed in Section 5 is applied to this dataset to find yearly periodicities of category 1 hurricanes across the eastern-pacific region. The following partial yearly periodicities have been detected (Table 2)-

Table 2  Yearly periodicities  of  category 1 hurricanes  in  the  eastern-pacific  region

| Timestamp(s) | Maximum certainty value reached across  this span ( in % ) |
|---|---|
| 30th  August  to  6th  September | 30 |
| 17th  September  to  21st  September | 24 |

Working with a time-series: The dataset contains daily average temperatures of Paris. The dataset is available at http://www.engr.udayton.edu/weather/. We  are  interested  in  detecting    if there is  any  yearly  periodicity  in  the  occurrence  of  a 10°F  temperature  rise  in  Paris. The  classic DTW technique described in Section 3 is used to first extract from the temperature time-series, all the time-intervals in which a 10°F temperature rise has occurred. Now, to find if there is any yearly periodicity in the occurrence of such a temperature rise, the technique described  in  Section 5  is  applied  to  the  time-intervals  that  were  extracted  from  the  time-series  by  the DTW  program. The  following  partial yearly periodicities  have been detected (Table 3)-

Table 3  Yearly  periodicities of  a  10°F temperature  rise  in  Paris

| Timestamp(s) | Maximum certainty value reached  across  this span( in % ) |
|---|---|
| 21st  April  to  27th  April | 50 |

## 8. CONCLUSION AND LINES FOR FUTURE WORK

A method  was  proposed   to  extract  calendar-based  periodicities  of   an  interval-based temporal  pattern i.e. a  pattern  that  occurs  across  a  sequence  of  time-intervals  in  either  a discrete  or  in  a  continuous  domain. For  this,  a  function  called  occurrence  function was defined  for  a  timestamp. A generalized  algorithm  was  developed  for  computing  the occurrence  function  at  any  timestamp  in  either  a  discrete  or  in  a  continuous  domain. Another  algorithm  for  finding  local  maxima  of  the  occurrence  function  was  also developed. It  was  shown  how  these  two  algorithms  could  be  used  to  determine  calendar-based  periodicities  of   an  interval-based  temporal  pattern  in  either  a discrete  or  a continuous  domain. The  extraction  of   periodicities   takes  O(n log n)  for  a  continuous domain  and  only  O(n)  for  a  discrete  domain (for hierarchical timestamps), where n is the number of intervals in which  the  pattern  occurs. The  proposed  technique  for  extracting calendar-based  periodicities  is  able  to  detect  both  partial  as  well  as  full  periodicities  of an interval-based temporal pattern with the same efficiency. Finally, a relationship between the periodicities of patterns  at  different  levels  of  a  time-hierarchy  was  also  presented.

Future  works  include  mining  of  causal relationships  and  correlations  among  multiple interval-based  temporal  patterns.

**Authors**

Mala Dutta  received the master's degree in Computer Applications from
Dibrugarh University in 2004. She has worked as a corporate trainer in the
Education and Research unit of Infosys Technologies Ltd., Bangalore. She is
currently a Ph.D. student in the Department of Computer Science at Gauhati
University. Her research interests focuses on temporal data mining and
pattern recognition.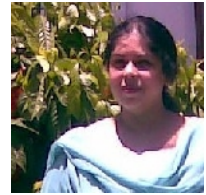

 Dr(Mrs). Anjana Kakoti Mahanta is currently holding the post of Professor in
the department of Computer Science, Gauhati University. She received the
master's degree in Mathematics from Gauhati University in 1986. She obtained
a Ph.D. degree in Computer Science from the same university in the year 1990.
In her Ph.D research, she worked in the area of Combinatorial Optimization.
Her current area of interest is Algorithms and Data Mining. Dr. Mahanta visited
the University of Warsaw for three months in the year 2007 under a bilateral
exchange program of Indian National Science Academy and Polish Academy of
Sciences. She has a good number of publications in journals and conference
proceedings at the national  and  international  level.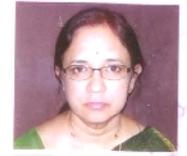